\documentclass{article}
\usepackage[utf8]{inputenc}
\usepackage[T1]{fontenc}
\usepackage{graphicx}
\usepackage{color}
\usepackage{cprotect}
\usepackage{url}
\usepackage{mathtools}
\usepackage{amssymb}
\usepackage{bbm}
\usepackage{float}

\usepackage{fullpage}

\title{Modelling the Impact of Scandals: the case of the 2017 French Presidential Election}
\author{Yassine Bouachrine and Carole Adam}
\date{\small{This is an ENSIMAG internship report originally written in June 2019 by intern Yassine Bouachrine under the supervision of Carole Adam}}

\begin{document}

\maketitle

\begin{abstract}
    This paper proposes an agent-based simulation of a presidential election, inspired by the French 2017 presidential election. The simulation is based on data extracted from polls, media coverage, and Twitter. The main contribution is to consider the impact of scandals and media bashing on the result of the election. In particular, it is shown that scandals can lead to higher abstention at the election, as voters have no relevant candidate left to vote for. The simulation is implemented in Unity 3D and is available to play online.
\end{abstract}

\section{Introduction}

During the 2017 French presidential election, the media had a very impactful role in the shift of the opinion away from the election's favorite François Fillon.
The seriousness of the accusations against the candidate led to Fillon plummeting in the polls.
We will try to model the impact of both conventional and social media through scandal diffusion, in order to better understand the dynamics underlying the voting process.

There are a variety of existing models for the voting process. However, most of the models see voting as a discrete event, while it is actually a sample at a given instant of a system in continuous evolution. This paper tries to shed some light on improvements which could take into account the inherent dynamism that comes with the interactions of the voters.

Another issue is that most computer voting models are adapted to the American context. In France, there are multiple candidates participating in the first round of the elections but in reality, few of them are actually considered by the voters, despite being aligned with their ideals. 
This is due to the nature of the voting process in France which takes place in two rounds. This leads the voters to cast their vote strategically towards candidates who have a chance of making it out of the first round.

The \textit{Voter Autrement}\footnote{Voter Autrement : \url{https://vote.imag.fr/}} experiment explored the effects of this strategic voting in 2017 by testing various voting systems during the presidential election.
The results showed that the alternative voting methods yield vastly different results (in terms of who is elected), especially for candidates such as J-L. Mélenchon and H. Hamon (systematic improvement) or F. Fillon and M. Le Pen (systematic decline). Other voting methods such as candidate ranking make the strategic vote useless, 
since every candidate gets a chance to make it to the second turn.

Finally, the context for the 2017 presidential election is even more unique in regard to the state of the French political scene. It is the first presidential election since the split of the centre-right party (UMP), and it is taking place amid the overall dissatisfaction of the population with the French Socialist Party (PS, left wing). This will be further elaborated on in the \textbf{Voting models} section as it is relevant to studying the impact of such a context on the existing models.

Our goal here is to model the impact of scandals over the course of the election and the change in the opinion of the voters.
Last year, A. Soutif \cite{albin2017}
tried to model the election process using an agent-based model by feeding the agents the results of the polls reported by the media. 
His goal in that study was to model the impact of the polls on the votes and on the strategic vote in particular, as polls give information on which candidates have chances of making it out of the first round.
Our approach aims to complement this work by showing the additional impact of media through the diffusion of scandals during the campaign. 

The first part of the paper (Section~\ref{sec:data}) focuses on the data analysis upon which the model is built. The second part of the paper (Section~\ref{sec:model}) addresses voting models and our implementation of the suggested model.

\section{Data analysis} \label{sec:data}

Building a model requires the availability of sufficient and relevant data about the phenomenon we are trying to model.

\subsection{Comparing poll results and media trends}

\begin{figure}[hbt]
    \centering
    \includegraphics[width=0.9\linewidth]{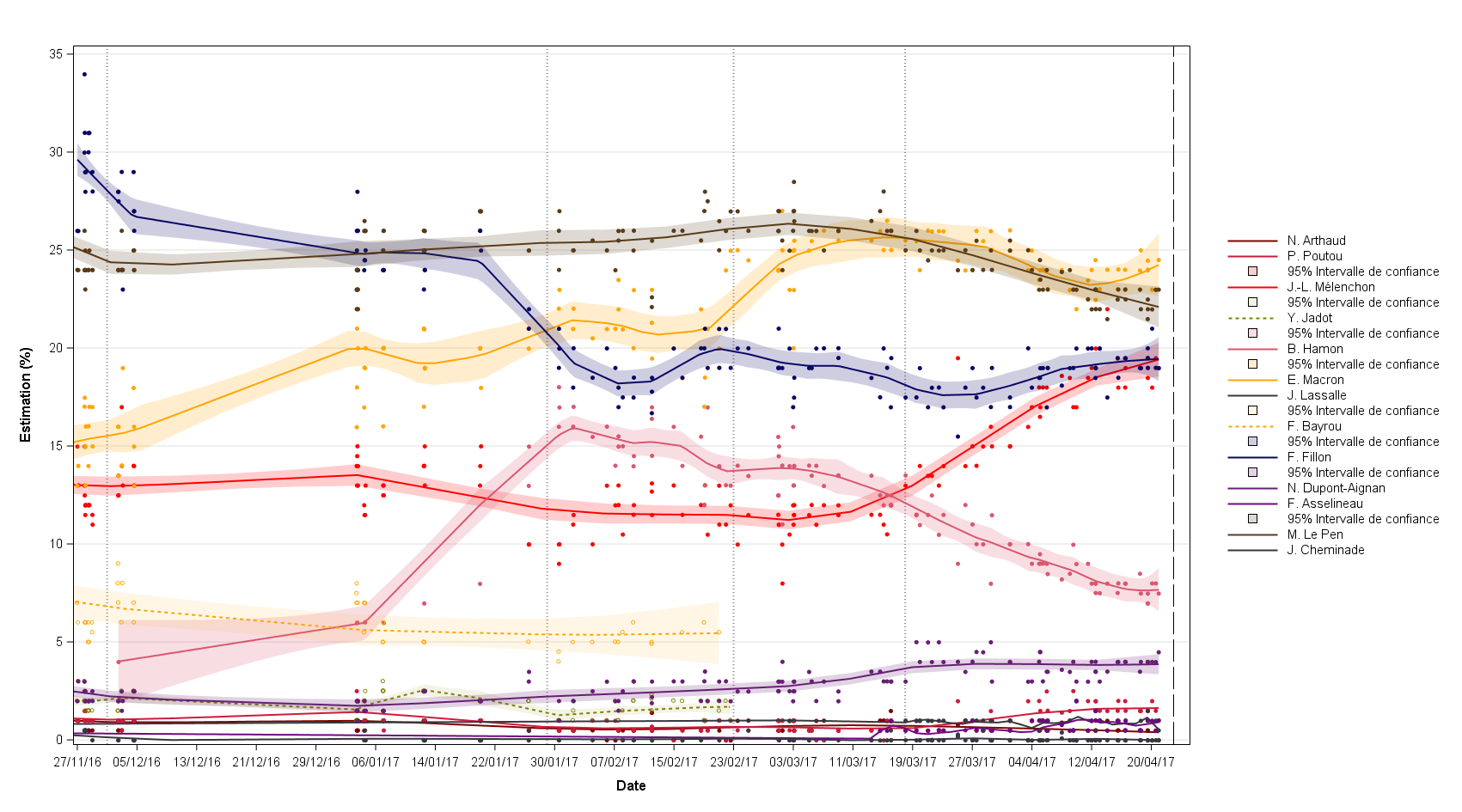}
    \caption{Evolution of the polls during the elections (Source: Wikipedia \cite{pollimg2017})}
    \label{fig:evolpolls}
\end{figure}

In our case, we took the aggregated poll results from various organisms \cite{polls2017} shown in Figure~\ref{fig:evolpolls},
and interpolated linearly when we had missing data (mostly B. Hamon in the early polls). We then compared the result of these polls with the evolution of the presence of the candidates across traditional media and Twitter.

\begin{figure}[hbt]
    \centering
    \includegraphics[width=0.9\linewidth]{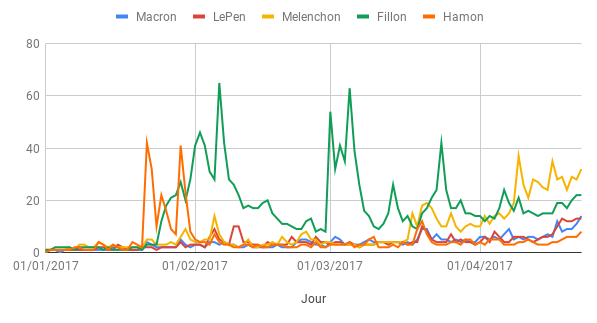}
    \caption{Evolution of the search queries for the candidates} \label{fig:trends}
\end{figure}

Google Trends\footnote{\url{https://trends.google.fr/trends/}} is a Google website that offers statistics on Google Search queries. We used it to evaluate the searched queries associated with the the top five candidates in the News section, the results are shown on Figure~\ref{fig:trends}. An interesting point can be made around how F. Fillon dominates the media presence in the early months of the election, alongside B. Hamon which we can ignore since it is mostly due to the French Socialist Party presidential primary. 

\begin{figure}[H]
    \centering
    \includegraphics[width=0.7\linewidth]{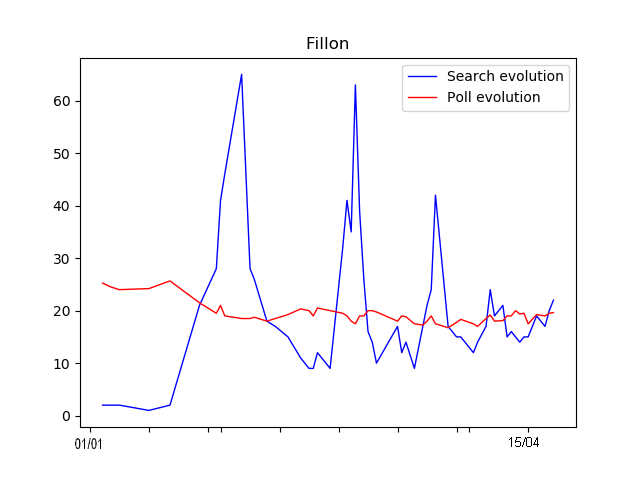}
    \caption{Evolution of the polls and search queries for Fillon}
    \label{fig:evolfillon}
\end{figure}

We can see that media coverage of this candidate spikes at multiple points in time (Figure~\ref{fig:evolfillon}), which is likely due to the ``Penelope gate'', which is a scandal associated with the alleged fictitious employment of members of Fillon's family. Further, the media frenzy surrounding the ``Penelope gate'' looks correlated with the evolution of the polls, as shown in Figure~\ref{fig:evolfillon}.

The takeaway from this comparison is that further media bashing does not seem to impact polls as much. This can be attributed to the nature of the scandal being a punctual event, and that people willing to cast away their vote already did so with the initial media outlets. The biggest beneficiary of this is definitely the closest candidate in the political spectrum: E. Macron.

\begin{figure}[hbt]
    \centering
    \includegraphics[width=0.7\linewidth]{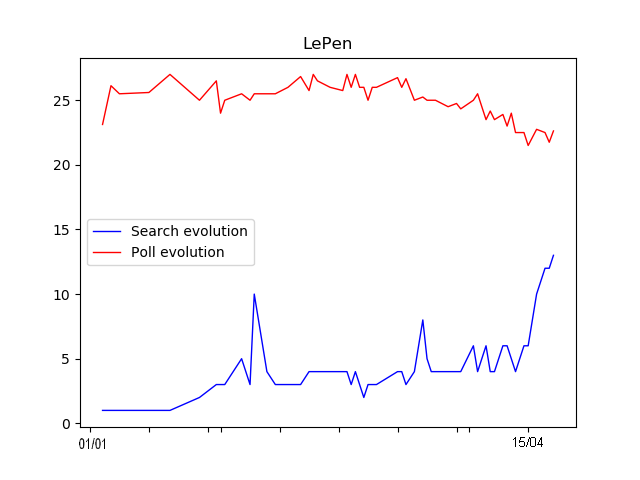}
    \caption{Evolution of the polls and search queries for Le Pen}
    \label{fig:evollepen}
\end{figure}

\newpage
Another interesting candidate is M. Le Pen who dropped significantly in the polls (Figure~\ref{fig:evollepen}) once the media picked up that she could actually finish first at the first round.

An hypothesis could be that there is actually a bidirectional relationship between polls and media coverage, with each one affecting the other. For example, in F. Fillon's case it is the disclosure of the scandal by \textit{Le Canard Enchaîné} that led to him dropping significantly, whereas in M. Le Pen's case, it is the poll results that led to an increase in coverage, and then her poll results dropped as a result of this increased coverage.

\subsection{Processing Twitter data}

The limits of the information we can leverage from Google Trends is that it does not tell us about the \textit{nature} or content of the coverage. We cannot know for sure if the increasing number of articles are rather positive or negative ones. We can deduce that \textit{a posteriori} by observing the impact on the polls, but it is what we are actually trying to model. Therefore, we used Twitter data, and analyze the tweets during the primary round to try and visualize opinion trends during the election.

\subsubsection{Data sets}

Two datasets were used in our work:
\begin{itemize}
    \item The first one is from \textbf{Kaggle} \footnote{\url{https://www.kaggle.com/jeanmidev/french-presidential-election}}: Kaggle is an online exchange platform for datascientist, users can publish datasets among other things. This Kaggle dataset contains tweets sampled during the elections. It is very rich but the data collection rate varies and some tweets appear to be truncated. More details about this dataset can be found at the source.
    \item The other dataset is a courtesy of E. Duble, research engineer at LIG. It contains an anonymized collection, sampling only geotagged tweets, that are mentioning the top hashtags during the election. The importance of hashtags has been shown for instance by the Politoscope project \cite{politoscope2017} 
\end{itemize}

\subsubsection{Clustering}
Building a vectorial representation of the tweets can be done in various ways. M. Campr and K. Jezek \cite{campr2015comparing} provided a performance evaluation of various methods for paragraph vectorization. 
At first, we opted for Tweet2Vec \cite{tweet2vec}, which relies on character-based representations (as opposed to word representations for Doc2Vec \cite{doc2vec}) that perform better for content such as tweets. However, Tweet2vec uses hashtag prediction to train the model, which is limited for our use-case since we already have a restricted number of hashtags. It also takes longer to train compared to word-based models.

We used Facebook fastText \cprotect\footnote{Facebook FastText: \url{https://fasttext.cc/}} to generate embeddings, and enhanced them with their term frequency-inverse document frequency (TF-IDF, \cite{tfidf}) in the corpus. We then computed the tweet embeddings as the average of the word embeddings.

Performing Principal Component Analysis (PCA, \cite{pca}) on the 100-dimensional tweet embeddings did not yield very good results as the embeddings are already built to minimize colinearity.

\section{Voting model} \label{sec:model} 

In this section, we provide an overview of existing models and their limitations before presenting the model we built through the observation of the data.

\subsection{Existing models and their limitations}

Doing a taxonomy of existing models is outside the scope of this paper, there are good resources available for that \cite{antunes2010theoretical, lewis2007economic, voting}. 
We will focus on some limitations of the existing models, starting with the psychological model. As hinted to in the introduction, the context of the election makes partisan identification hard to rely on: the schism of the centre-right party reshaped the political scene entirely. Also, the overall dissatisfaction with F. Hollande hurt the socialist party. Towards the end of his mandate, his popularity rating was lower than Macron's was during the Gilet Jaunes protests \cite{hollande2014}. 
On top of that, the appearance of new actors on the scene such as En Marche further shook the scene. En Marche made retrospective voting irrelevant as the party had never held responsibilities. The novelty Macron brings to the table, and his ambition of uniting the political parties, gave him a considerable advantage.

All of these circumstances made the elections very volatile. Even more sophisticated models such as the funnel of causality have to be rethought.
\begin{figure}[hbt]
    \centering
    \includegraphics[width=0.5\linewidth]{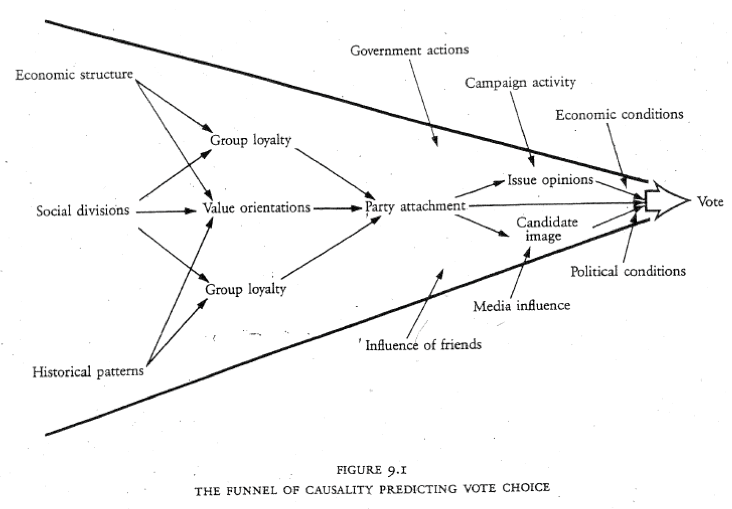}
    \caption{Funnel of causality, source \cite{dalton1988citizen}}
    \label{fig:funnel}
\end{figure}
Media has to have a bigger role in this funnel, especially social media as it has been shown to be a good indicator of standings \cite{anuta2017election},
almost as good as traditional polls. And that is, despite it being sensitive to social engineering (cf. Cambridge Analytica's impact on the American presidential election).

\subsection{Proposed model}

Our model is an enhancement of proximity models, in order to take into account the diffusion of scandals and the movement of neighboring agents.

\subsubsection{Simulation environment and initialization}

The environment is a 100 X 100 units 2D plan. There are two types of active agents: candidates and voters. 

For the simulation, we define:
\begin{itemize}
    \item The \textbf{appeasement delta} $\Delta\alpha \in [0,1]$, rate at which the repulsion of the candidate diminishes
    \item The \textbf{falloff rate} for the potential of the scandals $\Delta\rho \in [0,1]$
    \item The \textbf{maximum openness} for the voters $\sigma_{max} \in [0,100]$, defines how far a voter considers his surroundings  
    \item The \textbf{maximum tolerance} for the voters $\theta_{max} \in [0,+\infty[$
\end{itemize}

\subsubsection{Agents and their attributes}
For candidates $C$, we define: 
\begin{itemize}
    \item \textbf{Position} at time $t$ as $\psi_t \in [0,1]^2$ initialized manually
    \item \textbf{Repulsion} at time $t$ as $\gamma_t \in [0,1]$ with $\gamma_0 = 0$
    \item A list S of \textbf{scandals} with $S_i$ being the i-th one.
\end{itemize}
For voters $V$, we define: 
\begin{itemize}
    \item \textbf{Position} at time $t$ as $\psi_t \in [0,1]^2$ with $\psi_0 \sim \mathcal{U}^2 (0,1)$ (distance to a candidate inversely proportional to agreement with this candidate)
    \item \textbf{Openness} as $\sigma \in [0,\sigma_{max}]$ with $\sigma \sim clamp_{[0,1]}(\mathcal{N} (0.5,0.2^2))  \sigma_{max}$ (the radius is which a voter considers agents around him)
    \item \textbf{Charisma} as $\kappa \in [0,1]$ with $\kappa \sim clamp_{[0,1]}(\mathcal{N} (0.5,0.2^2))$ (the chance to influence others around)
    \item \textbf{Tolerance} as $\theta \in [0,1]$ with $\theta \sim clamp_{[0,1]}(\mathcal{N} (0.5,0.2^2)) \theta_{max}$ (the threshold for repulsion before dismissing a candidate completely)
    \item \textbf{Conformity} as $\eta \in [0,1]$ with $\eta \sim clamp_{[0,1]}(\mathcal{N} (0.5,0.2^2))$
\end{itemize}
For scandals, we define:
\begin{itemize}
    \item \textbf{Potential} at time $t$ as $\rho_t \in [0,1]$ with $\rho_0$ initialized manually by the user (parameter in the simulation)
\end{itemize}

\subsubsection{Simulation update}

At each time-step of the simulation, we update the entities. To simplify the equations, we assume that the values are clamped to their domain.

For scandals, the potential decreases with time, at the falloff rate:
\begin{equation}
    \rho_{t+1}(x) = \rho_{t}(x) - \Delta\rho
\end{equation}
For candidates, the position is static, and the repulsion increases with each scandal, and decreases with time at the pace set by the appeasement delta:
\begin{equation}
    \gamma_{t+1}(x) = \gamma_{t}(x) - \Delta\alpha + \sum_{y \in S(x)} \rho_{t+1}(y)
\end{equation}
For voters, the position evolves as their opinion about the different candidates evolves based on their surroundings:
\begin{equation}
\begin{split}
    \psi_{t+1}(x) & = \psi_{t}(x) \\
    & + \operatorname*{argmin}_y \{||\psi_t(x) - \psi_t(y)||_2 \mid y \in C \land \gamma_{t+1}(y) < \theta(x)\} \frac{1}{1 + \gamma_{t+1}(y)}  \\
    & + \eta\sum_{y \in V \land ||\psi_t(x) - \psi_t(y)||_2 < \sigma(x)}\kappa(y)(\psi_t(x) - \psi_t(y))
\end{split}
\end{equation}
When the simulation stops, each voter votes for the closest candidate still considered in the openness radius around him. If there are none, the voter withholds his vote (chooses abstention).


\subsubsection{Implementation details}

The simulation is built in Unity 2018.3.5f1 \footnote{Unity 3D: \url{https://unity.com/}}. It allowed for faster prototyping and also supports a wide range of platforms to run the simulation on.
The simulation is available to play online at \url{http://lig-tdcge.imag.fr/votsim/} or to download as a WebGL export\footnote{WebGL export of the simulator: \url{https://ensiwiki.ensimag.fr/index.php?title=IRL_-_Modélisation_de_la_dynamique_des_opinions_des_électeurs}}.

\begin{figure}[hbt]
    \centering
    \includegraphics[width=0.5\linewidth]{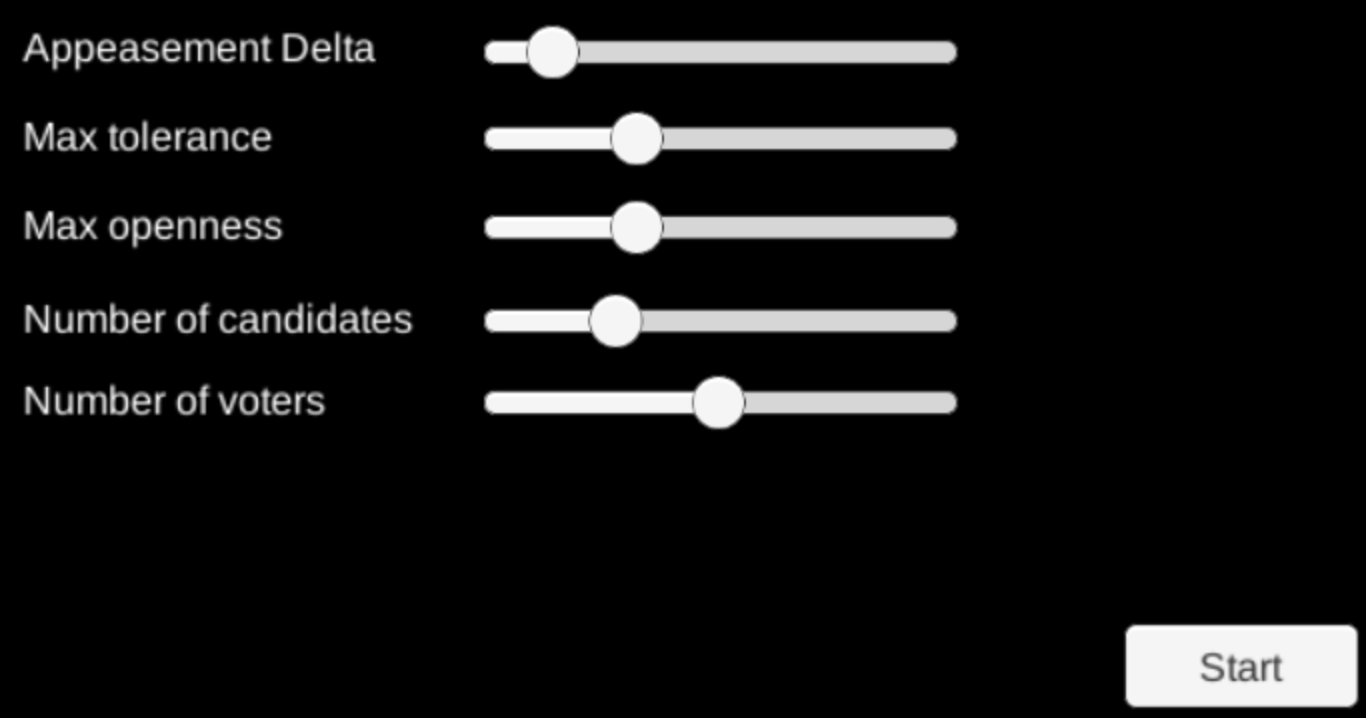}
    \caption{Parameter selection before starting the simulation}
    \label{fig:params}
\end{figure}

A first screen lets the user select the values of the global parameters of the simulation (Figure~\ref{fig:params}): the number of voters and candidates, the appeasement delta (rate at which the scandals decrease, which determines the duration of their effect on opinions), and the maximum values for tolerance and openness (individual values of all agents are then set randomly under this boundary).

\begin{figure}[hbt]
    \centering
    \includegraphics[width=0.8\linewidth]{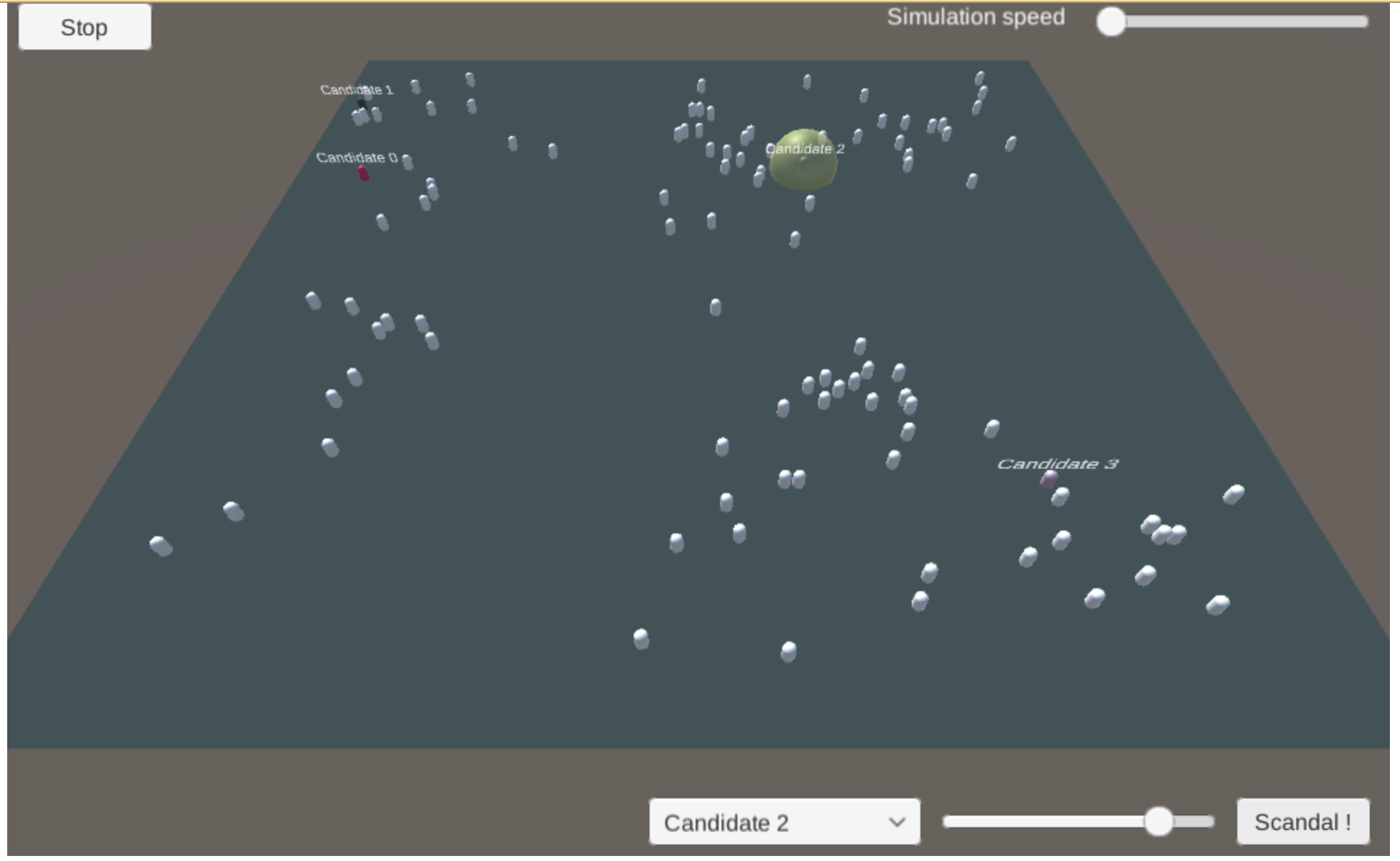}
    \caption{Screenshot of the simulation after triggering a scandal}
    \label{fig:screenshot}
\end{figure}

On the next screen, the user can modify the simulation speed. Voters are moving in the environment towards or away from the candidates. The user can also trigger a scandal and choose its intensity and target candidate, by using a button at the bottom of the window, in order to observe the influence on the movements of the voters (see Figure~\ref{fig:screenshot}). The intensity of the scandal then decreases with time. 

\subsubsection{Discussion and results}

Regarding the motivations behind the model, we wanted to enhance the existing models with the observations made from our data analysis. First, regarding the initialization, the justification behind the uniform distribution for the voters' positions is the unique context behind the 2017 election that we detailed earlier, with many voters not knowing which parties to consider. A fine-tuned Gaussian mixture model could also be explored.

Most of the reasoning behind the model is based on the reactions to the Penelopegate, with some voters completely turning their backs on F. Fillon (which we model by a scandal being above their tolerance threshold) and some only showing hesitation (tolerance threshold not reached). A temporary dip followed by a partial recovery in the polls supports the model: regardless of the further media bashing around the event, voters have a threshold over which further coverage has no effect.

Over the simulated scenarios, one of the most interesting observations is that scandals tend to be tied with an increased abstention rate. In our model this is represented by the voter moving too far from all candidates (due to repulsion generated by scandals about their favourite candidates, or to diverging opinions with the others), so that no valid candidate is still present in the openness radius when the election comes; in that case the voter prefers to choose abstention. Our model can therefore reproduce and explain the 2017 presidential election's high abstention rate in the first round, at 22.23\% \cite{abstention2016}.


\section{Conclusion}  \label{sec:conclu}

We have seen that media coverage of the campaign scandals can have a big impact on the election results. The simulation showed that scandals can totally shape the result of the election and that scandals profit to the closer candidates on the political spectrum. The more interesting finding was how scandals impact the abstention rate, which is in agreement with the observations made in the context of the French 2017 presidential election and the high abstention rate recorded. 


There is still much to do to reach a unified model, a first step in that direction would be enhancing the simulation with the results from A. Soutif's experiments regarding the impact of the polls \cite{albin2017}.
We could then initialize the model to match the French political scene at the beginning of the first round and test if it corresponds to the observed election results.
If the model is validated, we could explore alternative scenarios for the election: how different scandals could have led to different results and particularly what would have happened if there were no scandals involving the pre-campaign favorite F. Fillon.

\footnotesize

\end{document}